# Cross-Sectional Scanning Tunneling Microscopy and Spectroscopy of Semimetallic ErAs Nanostructures Embedded in GaAs

Jason K. Kawasaki[a]

University of California Santa Barbara, Materials Department, Santa Barbara CA 93106

Rainer Timm

Lund University, Nanometer Structure Consortium (nmC@LU), Department of Physics, Sweden

Trevor E. Buehl

University of California Santa Barbara, Materials Department, Santa Barbara CA 93106

Edvin Lundgren, Anders Mikkelsen

Lund University, Nanometer Structure Consortium (nmC@LU), Department of Physics, Sweden

Arthur C. Gossard and Chris J. Palmstrøm[b]

University of California Santa Barbara, Materials Department and Electrical and Computer Engineering Department, Santa Barbara CA 93106

[a]Electronic Mail: jkawasaki@engineering.ucsb.edu

[b]American Vacuum Society member

[c]This article is based on material presented at the 27th North American Molecular Beam Epitaxy Conference

The growth and atomic/electronic structure of molecular beam epitaxy (MBE)-grown ErAs nanoparticles and nanorods embedded within a GaAs matrix are examined for the first time via cross-sectional scanning tunneling microscopy (XSTM) and spectroscopy (XSTS). Cross sections enable the interrogation of the internal structure and are well suited for studying embedded nanostructures. The early stages of embedded ErAs nanostructure growth are examined via these techniques and compared with previous cross sectional TEM work. Tunneling spectroscopy *I(V)* for both ErAs nanoparticles and nanorods was also performed, demonstrating that both nanostructures are semimetallic.

## I.  INTRODUCTION

Epitaxial rare-earth-monopnictide (RE-V) nanostructures embedded within a semiconductor matrix have garnered great attention due to a number of exciting properties, including surface plasmon resonances,[1] phonon scattering for high figure of merit thermoelectrics,[2] ultrafast recombination for terahertz devices,[3] and tunneling enhancement.[4] Among such structures, ErAs nanoparticles embedded within GaAs are one of the most widely studied.

ErAs has cubic rocksalt structure ($a_{ErAs}$ = 5.73 Å) and grows epitaxially[5,6] within zincblende GaAs ($a_{GaAs}$ = 5.65 Å) with a continuous arsenic sublattice across the ErAs/GaAs interfaces.[7-9] On the GaAs (001) surface, ErAs forms via an embedding mechanism in which Er exchanges with Ga and the Ga is ejected from the surface.[10] Due to the limited diffusivity of Er in ErAs, the particles typically embed themselves no deeper than 3-4 ML into the surface before extending laterally in the (001) plane.[10,11]

In contrast, when grown by codeposition with GaAs on high index surfaces, e.g. GaAs (114)A, ErAs has been shown to form ordered arrays of nanorods instead of nanoparticles.[12] At a composition of 6 at.% Er, these nanorods align axially along [112] with spacings of roughly 7 nm and diameters of roughly 2 nm. Like the nanoparticles on (001) substrates, the nanorods on (114)A have a continuous arsenic sublattice across the ErAs/GaAs interfaces. However, due to their anisotropic nature, they also show interesting optical properties under polarized light and show great promise for embedded nanowire devices.[12] Additionally their mechanisms for growth and ordering are still largely unknown.[12]

Both ErAs nanoparticles and nanorods have previously been studied via transmission electron microscopy (TEM),[8,12,13] but to date only few plan view scanning tunneling microscopy (STM) studies have been made on ErAs nanoparticles,[10] and no cross sectional STM studies have been made for either system. Furthermore there exists some controversy regarding the electronic structure of embedded ErAs nanostructures. In particular, it is unclear whether ErAs nanoparticles are semimetallic as in bulk or whether they become semiconducting due to quantum confinement.[14]

Here we examine the early stages of growth and the electronic structure of molecular beam epitaxy (MBE)-grown ErAs nanostructures embedded within GaAs. Our characterization method is a combination of cross-sectional scanning tunneling microscopy (STM) and spectroscopy (STS), which enables the investigation of internal structure and is particularly suited for characterizing embedded nanostructures.[15,16] Two systems have been studied: nanoparticles embedded within a GaAs matrix grown on (001) GaAs, and nanorods embedded within a GaAs matrix grown on (114)A GaAs. The nanoparticles on (001) GaAs were grown at varying coverages of ErAs in order to study the initial stages of nucleation and growth. Tunneling point spectroscopy measurements were used to determine the electronic properties of the nanoparticles and nanorods.

## II. EXPERIMENTAL

Both the nanoparticles and nanorods were grown by molecular beam epitaxy (MBE) from effusion cell sources. The nanoparticle samples were grown on n-type GaAs

(001) substrates in a modified VG V80H system with a base pressure $\leq 5\times10^{-11}$ Torr. The epitaxial layer sequence consisted of GaAs (001) substrate / 170 nm GaAs / 50 × (20 ML AlAs/20 ML GaAs superlattice) / 500 nm GaAs / 4 x (*X* ML ErAs/125 nm GaAs) / 500 nm GaAs, where *X* = 0.125, 0.25, 0.5, 1.0 (Fig. 1). Layers were grown at rates of 0.5 ML/s for GaAs and AlAs and 0.025 ML/s for ErAs. Growth rates were determined by RHEED oscillations on calibration samples prior to growth of the STM samples.[17] All layers were grown under an $As_2$ overpressure with an effective As:Ga incorporation ratio of roughly 6:1 (determined using As and Ga RHEED oscillations during GaAs growth on a calibration sample), a constant substrate temperature of 540 ˚C, and constant Si doping of ~$5\times10^{18}$ cm$^{-3}$.

The nanorod samples were grown on n-type gallium-polar (114)A GaAs substrates in a Varian Gen II system. The epitaxial layer sequence consisted of GaAs (114)A substrate / 150 nm GaAs / 50 × (20 ML AlAs/20 ML GaAs superlattice) / 1000 nm GaAs / 500 nm ErAs:GaAs / 10 nm GaAs. The ErAs:GaAs layer was grown by codeposition of Er, Ga, and As, with a 6% atomic concentration of ErAs and a substrate temperature of 580 ˚C. All layers had a Si doping concentration of roughly $10^{18}$ cm$^{-3}$. Further growth details are provided in a previous publication.[12]

After growth, the samples were mounted *ex situ* onto vertical STM cross-sectional sample holders and loaded into an Omicron variable temperature STM with base pressure $< 10^{-10}$ Torr. Samples were then outgassed at 200 ˚C for 2 hours and cleaved in UHV to expose a clean {110} surface. Tungsten STM tips were prepared by electrochemical etching and cleaned *in situ* by sputtering. STM images were acquired for sample biases in the range -1.8 to -2.5 V and tunneling currents of 80 to 120 pA.

STS point spectroscopy was performed by interrupting the feedback and measuring the tunneling current $I_m(V)$ at specified points on the {110} surface. In order to amplify the conductance signal and gain a greater dynamic range, the variable tip-sample separation procedure described by Feenstra was used.[18] Here the separation is given by $z(V) = z + \alpha |V|$ where $\alpha = 0.15$ nm/V.

## III. RESULTS AND DISCUSSION

Fig. 1 presents an overview of the STM images for ErAs nanoparticles embedded in GaAs (001). All scans are made on the {110} surface at negative sample bias (filled states). The AlAs/GaAs superlattice serves as a marker to determine the position of the subsequent ErAs layers. The 1-2 ML thin clusters observed in the GaAs spacers and AlAs/GaAs superlattice are Si precipitates due to heavy doping and are identical to those observed in STM by Domke et al.[19]

Fig. 2 shows high magnification STM images of the individual ErAs particles. The vertical lines are As atomic rows on the GaAs {110} surface. Since the {110} is not the rocksalt ErAs cleavage plane, the particles tend not to cleave. Instead, the particles remain stuck in one of the cleavage surfaces and are pulled out of the other. This results in protruding particles (Fig. 2a) or holes due to missing particles (Fig. 2b) in the cross sectional STM images. For coverages of 0.25 and 0.5 ML ErAs, the protruding particles appear nearly spherical with an average diameter of 2-3 nm. Missing particles also have dimensions of 2-3 nm, but while some missing sites appear spherical, others have nearly square cross sections with facets parallel to the [110] and [001] directions. This observation of faceting is consistent with TEM work by Kadow et al,[20] who observed that ErAs particles form facets in crystallographic directions, with edges oriented along [100]. We suspect that all of our particles are in fact faceted and that some particles simply appear more spherical because the particles crumble upon cleavage.

From height profiles of the missing ErAs particles, it should be possible in principle to determine the facet planes. Crystallography and Madelung energy considerations suggest that the energetically favored facets are the ErAs rocksalt {100} and the zinc blende GaAs {110}. If the interfaces are {100}-type one would expect height profiles along [001] to be triangular with 45° inclination and height profiles along [110] to be rectangular. However, for [001] height profiles of missing ErAs clusters, we observe both rectangular and triangular type profiles, with a maximum depth of roughly 2 Å (an example rectangular profile is shown in Fig. 2b, right). [110] height profiles also show both rectangular and triangular profiles with a maximum depth of 2 Å. These 2 Å depth profiles are surprisingly shallow, which suggests that either the ErAs is crumbling, or that the STM tip is not sharp enough to resolve the full depth of the missing particles. Additionally the variations in profile shape make it difficult to determine the preferred facet plans, and we attribute these variations due to crumbling of the particles. Finally, note that the 3-4 nm depth along [001] measured by cross sectional STM is larger than the expected 2 nm (4 ML) diffusion-limited depth,[10] a further indication that the particles may be crumbling.

A buried ErAs nanoparticle is shown in Fig. 2c. Here we see a smooth profile 0.7 Å in height overlayed on the atomic corrugation. This profile is Gaussian in shape with a standard deviation of $\sigma = 4.1$ nm and full width half maximum of 4.8 nm. The apparent height further reduces from 0.7 Å to 0.5 Å when the bias voltage is changed from -1.8 V to -2.0 V, and thus it represents an electronic feature rather than a topographical feature. This site is interpreted to be a buried ErAs particle whose electronic states induce electronic changes in the surrounding GaAs matrix, such as band bending or electronic states in the GaAs band gap. If we consider the effective diameter of electronic contrast to be roughly $2\sigma = 8.2$ nm, we find that this buried contrast is larger than the 3-4 nm diameter particles themselves.

Large area scans of the four ErAs layers of varying coverage are shown in the top panels of Fig. 1. We find that with increasing ErAs coverage from 0.125 to 1.0 ML, the density of ErAs particles within the GaAs matrix increases. In the 1.0 ML region a step 1 ML in height is produced upon cleavage, and this step is likely due to the increased particle density. It also appears that for low ErAs coverages in the range 0.125 to 0.5 ML the particles are nearly spherical (or cubic), with equal height and width in the [001] and [110] directions, and at a larger coverage of 1.0 ML the particles begin to elongate in the (001) plane. These results are consistent with TEM results by Driscoll et al[13] and plan view STM by Schultz et al,[10] who found that ErAs particles first embed themselves 3-4 monolayers in depth before extending laterally in the (001) plane.

Fig. 3 shows a plot of the particle density (number of particles per length along the layer) versus ErAs coverage. This plot was compiled by examining successive scan frames from each ErAs layer (100 nm frame height) and counting the number of particles in each frame. The number of particles N was then averaged over 5 succesive frames (500 nm total length) for the 0.125 ML layer, and 10 successive frames (1 μm total length) for the 0.25, 0.5, and 1.0 ML layers. The error bars represent the statistical error, defined as the square root of N per frame length. In the range 0.125 to 0.5 ML the plot is linear, in agreement with the observation that in the low coverage regime, the number of particles increases while the particle size stays roughly constant. For coverages larger than 0.5 ML the plot begins to break from linearity and curve downward. By conservation, here the excess ErAs must be contributing to increase the apparent particle size, and this is consistent with observations that at larger coverages, the particles begin to extend laterally in the (001) plane. Alternatively, the increase in apparent size could be due to coalesence of individual particles.

Cross-sectional STM studies were also performed on ErAs nanocomposites grown on (114)A GaAs surfaces under conditions that produce oriented embedded nanorods (Fig. 4). Like the nanoparticle samples, the nanorod samples are also scanned on a cleaved {110} surface. Here due to the difficulty in cleaving high index plane substrates, several cleavage steps are produced; however both the [110] GaAs atomic rows and ErAs nanocomposites are still resolved. Due to the rough cleave it is difficult to determine whether these nanocomposites are continuous nanorods or rows of oriented nanoparticles, but previous cross sectional TEM work has suggested that they are in fact continuous nanorods.[12] Based on a measured angle of 52 (±5)° between the nanorodsand the [110] atomic rows, the nanorods are found to be oriented along [112] (54.7° expected angle, in the {110} cleavage plane), in good agreement with TEM studies by Buehl et al.[12] The nanorods have an average diameter on the order of 2-3 nm, also in agreement with Buehl et al. Further studies are required to develop methods for cleaving and obtaining higher resolution STM images of the nanorod/GaAs interface.

Tunneling point spectroscopy was used to probe the electronic structure of the embedded ErAs nanostructures. Fig. 5 shows tunneling current $I_0(V)$ curves for the GaAs matrix, protruding ErAs nanoparticles, and protruding ErAs nanorods. Here we have compensated for the variable gap by multiplying the measured current $I_m(V)$ by an exponential, i.e. $I_0(V) = I_m(V)\exp(2\kappa z(V))$ where $z(V)$ is the variable tip-sample separation defined previously and we use a decay length of $\kappa = 1$ Å$^{-1}$.[18] The GaAs curve is averaged over 20 different spectra in the GaAs capping layer, the ErAs nanoparticle curve is averaged over spectra from 20 different protruding nanoparticles, and the nanorod curve is averaged over 10 spectra on 5 different nanorods. The GaAs $I_0(V)$ appears as expected from the literature, with zero tunneling current in the bandgap (roughly -0.9 to 0.7 V) and finite tunneling at larger negative and positive sample biases.[18] We also observe that the Fermi level (V = 0) is pinned near mid gap, which is likely due to the large concentration of Si precipitates/surface defects (for an example, see Fig. 1, AlAs/GaAs region).[21-23]

In contrast, the ErAs nanoparticle and nanorod $I_0(V)$ have finite tunneling through the GaAs band gap and have a finite slope through the Fermi level at V = 0, indicating that these ErAs nanostructures are semimetallic and not semiconducting. The slight leveling in slope at the Fermi level suggests that the nanorods and nanoparticles are semimetallic, as a metal should exhibit a linear $I_0(V)$. Finally, we note that spectra measured directly over buried particles (Fig. 2c) are nearly identical to $I_0(V)$ centered over protruding particles, and thus the observed semimetallic behavior is not induced by cleaveage defects and instead results from the particles themselves.

## IV. SUMMARY AND CONCLUSIONS

Cross sectional scanning tunneling microscopy and spectroscopy were performed on ErAs nanoparticles and nanorods embedded in a GaAs matrix. This method enables the atomic scale investigation of internal structure and is well suited for the study of embedded nanostructures. For ErAs nanoparticles embedded in (001) GaAs, it is found that during the initial stages of growth the particle density increases nearly linearly with ErAs coverage. With increasing ErAs coverage (0.5 to 1.0 ML) the particle density begins to dip below linearity, which marks the onset of particle elongation in the (001) plane. ErAs nanorods embedded in (114)A GaAs were found to order along [112], in good agreement with previous cross sectional TEM results.[12] Tunneling spectroscopy for both nanorods and nanoparticles showed a nonzero slope in $I_0(V)$ at the Fermi level, indicating semimetallic rather than semiconducting properties. Further spectroscopy studies are needed in order to examine the effects of quantum confinement on ErAs nanostructures and the formation of interface states and band bending at the ErAs/GaAs interfaces.[24]

# ACKNOWLEDGMENTS

This work was supported by the National Science Foundation through the IMI Program (Award No. DMR 0843934) and the UCSB MRL (Award No. DMR 05-20415), the Army Research Office (Award No. W911NF-07-1-0547), and Air Force Office of Scientific Research (Award No. FA9550-10-1-0119). Additionally, J.K. acknowledges funding from the Air Force Office of Scientific Research through the NDSEG Fellowship. This work was also supported by the Swedish Research Council (VR), the Swedish Foundation for Strategic Research (SSF), the Crafoord Foundation, the Knut and Alice Wallenberg Foundation and the European Research Council under the European Union's Seventh Framework Programme (FP7/2007-2013) / ERC Grant agreement n° [259141].

[1] M. P. Hanson, A. C. Gossard, and E. R. Brown, J. Appl. Phys. **102**, 1 (2007).

[2] W. Kim, S. L. Singer, A. Majumdar, D. Vashaee, Z. Bian, A. Shakouri, G. Zeng, J. E. Bowers, J. M. O. Zide, and A. C. Gossard, Appl. Phys. Lett. **88**, 242107 (2006).

[3] C. Kadow, S. B. Fleischer, J. P. Ibbetson, J. E. Bowers, A. C. Gossard, J. W. Dong, and C. J. Palmstrøm, Appl. Phys. Lett. **75**, 3548 (1999).

[4] J. M. O. Zide, A. Kleiman-Shwarsctein, N. C. Strandwitz, J. D. Zimmerman, T. Steenblock-Smith, A. C. Gossard, A. Forman, A. Ivanovskaya, and G. D. Stucky, Appl. Phys. Lett. **88**, 162103 (2006).

[5] I. Poole, K.E. Singer, and A.R. Peaker, J. Cryst. Growth **121**, 121 (1992).

[6] P. Rutter, K.E. Singer, and A.R. Peaker, J. Cryst. Growth **182**, 247 (1997).

[7] N. G. Stoffel, C. J. Palmstrøm, and B.J. Wilkens, Nuc. Instr. Meth. Phys. Res. B. **56-57**, 792 (1991).

[8] D. O. Klenov, J. M. O. Zide, J. D. Zimmerman, A. C. Gossard, and S. Stemmer, Appl. Phys. Lett. **86**, 241901 (2005).

[9] A. Guivarc'h, Y. Ballini, M. Minier, B. Guenais, G. Dupas, G. Ropars, and A. Regreny, J. Appl. Phys. **73**, 8221 (1993).

[10] B. D. Schultz and C. J. Palmstrøm, Phys. Rev. B. **73**, 1 (2006).

[11] C. J. Palmstrøm and T. D. Sands, in *Contacts To Semiconductors*, edited by L.J. Brillson (Noyes, 1993), p. 67.

[12] T. E. Buehl, J. M. LeBeau, S. Stemmer, M. A. Scarpulla, C. J. Palmstrøm, and A. C. Gossard, J. Cryst. Growth **312**, 2089 (2010).

[13] D. Driscoll, M.P. Hanson, E. Mueller, and A.C. Gossard, J. Cryst. Growth **251**, 243 (2003).

[14] M. A. Scarpulla, J. M. O. Zide, J. M. LeBeau, C. G. Van De Walle, A. C. Gossard, and K. T. Delaney, Appl. Phys. Lett. **92**, 173116 (2008).

[15] R. Timm, A. Lenz, H. Eisele, L. Ivanova, M. Dahne, G. Balakrishnan, D. L. Huffaker, I. Farrer, and D. A. Ritchie, J. Vac. Sci. Tech. B. **26**, 1492 (2008).

[16] A. Mikkelsen and E. Lundgren, Prog. Surf. Sci. **80**, 1 (2005).

[17] C.J. Palmstrom, S. Mounier, T.G. Finstad, and P.F. Miceli, Appl. Phys. Lett. **56**, 382 (1990).

[18] R. M. Feenstra, Phys. Rev. B. **50**, (1994).

[19] C. Domke, P. Ebert, M. Heinrich, and K. Urban, Phys. Rev. B. **54**, 10288 (1996).

[20] C. Kadow, J.A. Johnson, K. Kolstad, and A.C. Gossard, J. Vac. Sci. Tech. B. **21**, 29 (2003).

[21] M. D. Pashley and G. P. Srivastava, Phil. Trans. Royal Soc. A. **344**, 533 (1993).

[22] R.M. Feenstra and P Mårtensson, Phys. Rev. Lett. **61**, 227 (1988).

[23] R. Tung, Mater. Sci. Eng. R. **35**, 1 (2001).

[24]J. Kawasaki, R. Timm, T. E. Buehl, E. Lundgren, A. Mikkelsen, A. C. Gossard, and C. J. Palmstrøm, To Be Published (2011).

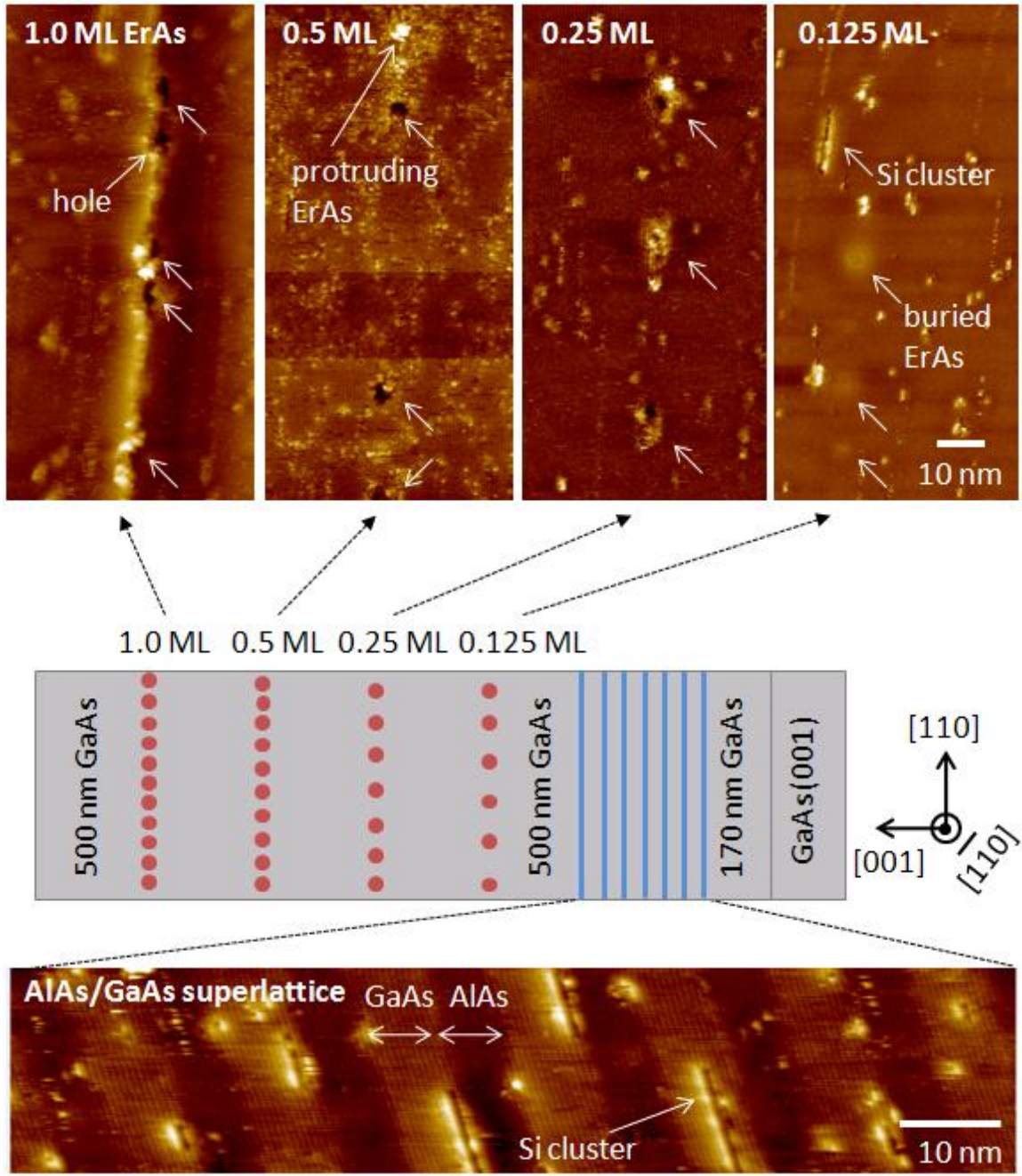

Figure 1. (Color online) Schematic sample structure and filled states STM images of the {110} cleavage plane. The sample consists of varying coverages of ErAs separated by 125 nm of GaAs on a GaAs (001) substrate. Depending on their location relative to the cleavage plane, the ErAs can appear as protruding particles, buried particles, or holes. The unlabelled arrows denote ErAs sites.

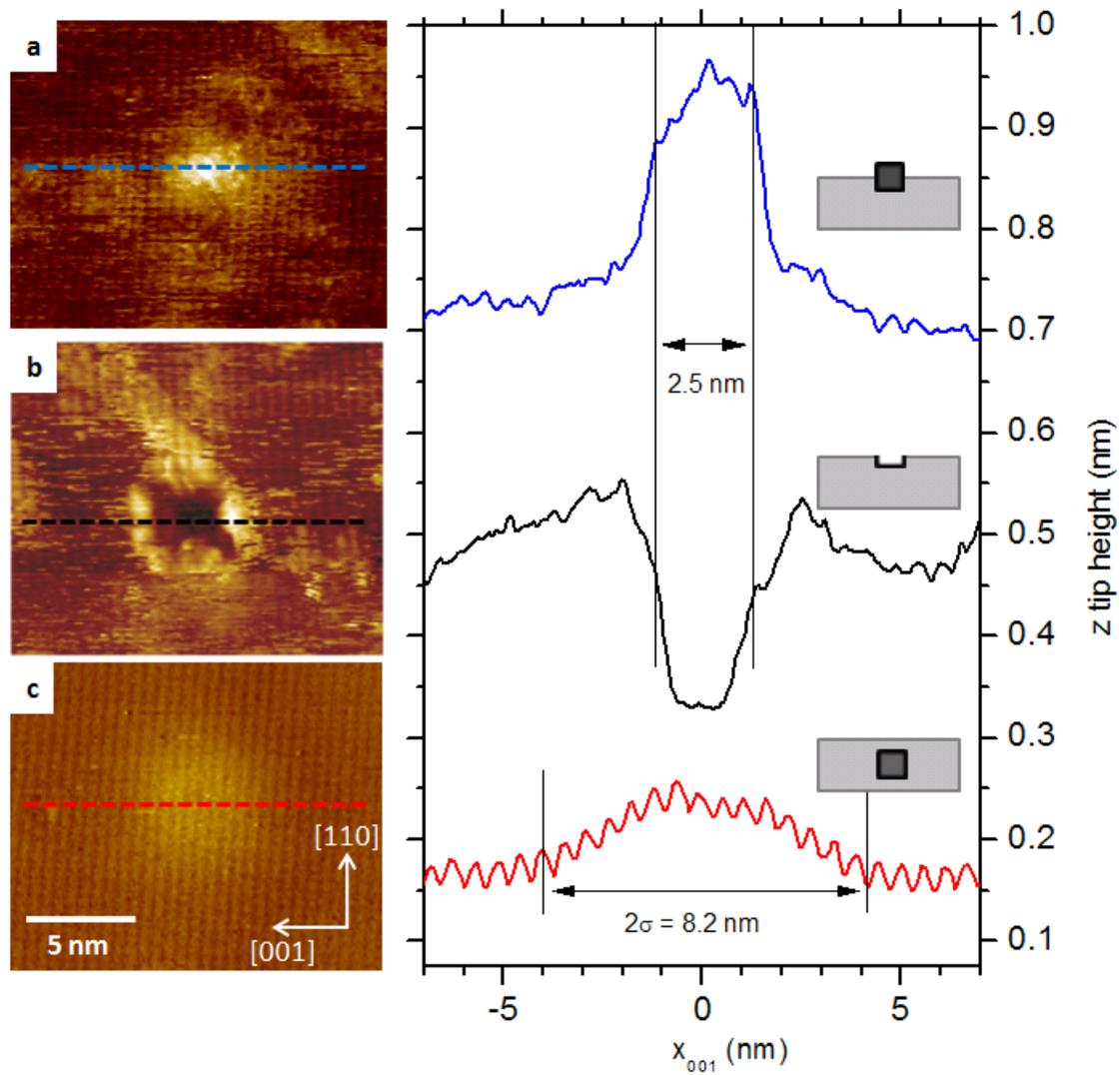

Figure 2. (Color online) Filled states STM images and height profiles of (a) protruding, (b) pulled-out, and (c) buried ErAs nanoparticles. (a) and (b) are from the region of 0.5 ML ErAs coverage. The apparent height of the protruding particle is 2 Å, and the apparent depth of the pulled-out particle is 2 Å. This particluar pulled-out particle had a nearly rectangular height profile. (c) is from the 0.125 ML coverage and the apparent height of the buried particle is 0.7 Å. Note the vertical lines are As atomic rows on the GaAs {110} surface. Insets: schematics of the three types of particle sites.

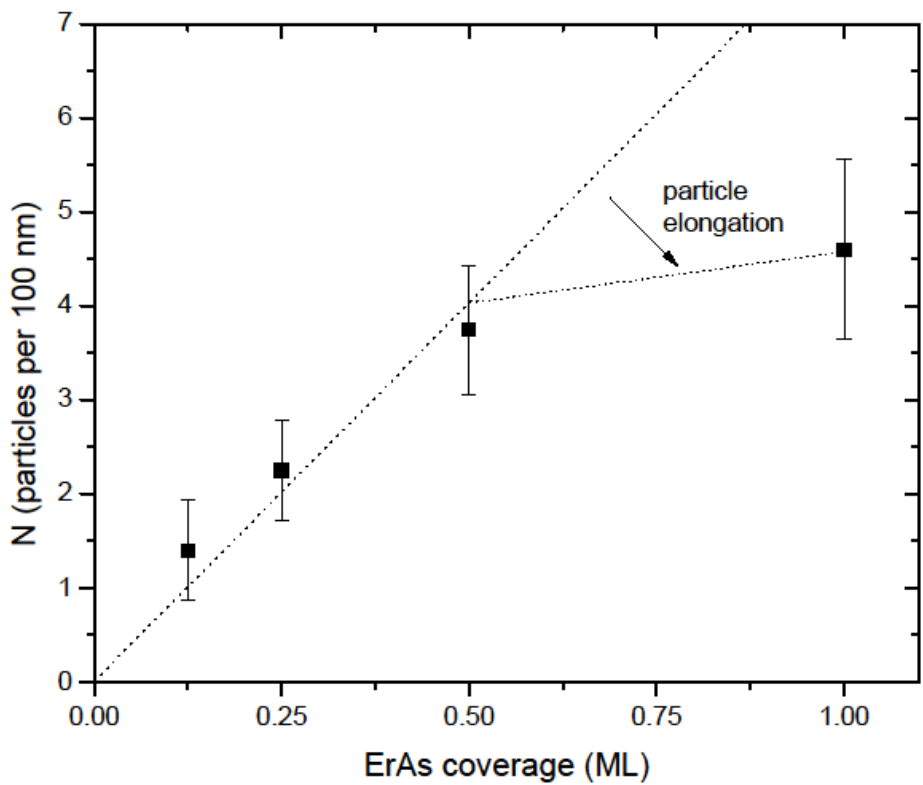

Figure 3. Plot of the particle density (number of particles per 100 nm) as a function of ErAs coverage. At low coverages (0.125 to 0.5 ML), the number of particles increases while the particle size remains roughly constant. The departure from linearity for coverages greater than 0.5 ML corresponds with the particles growing/elongating in the (001) plane.

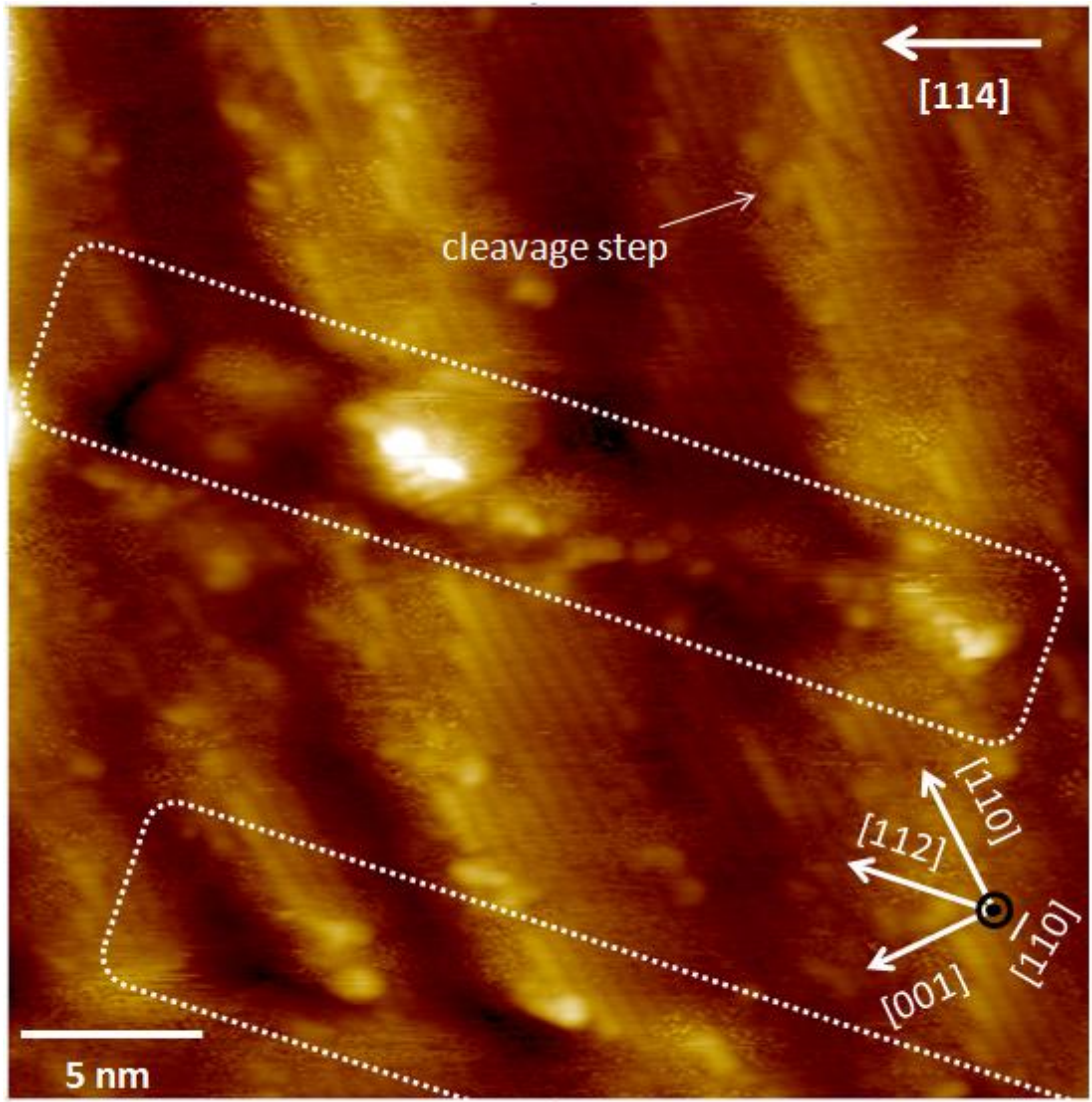

Figure 4. (Color online) Filled states STM images of ErAs nanocomposites grown on (114)A GaAs surfaces under conditions that produce oriented embedded nanorods. The rectangular boxes in the figure show the location of the ErAs nanorods. The nanorods align along [112] and are roughly 2 nm in diameter. The diagonal lines are As atomic rows aligned along [110]. Due to difficulties in cleaving high index plane substrates, several cleavage steps are produced.

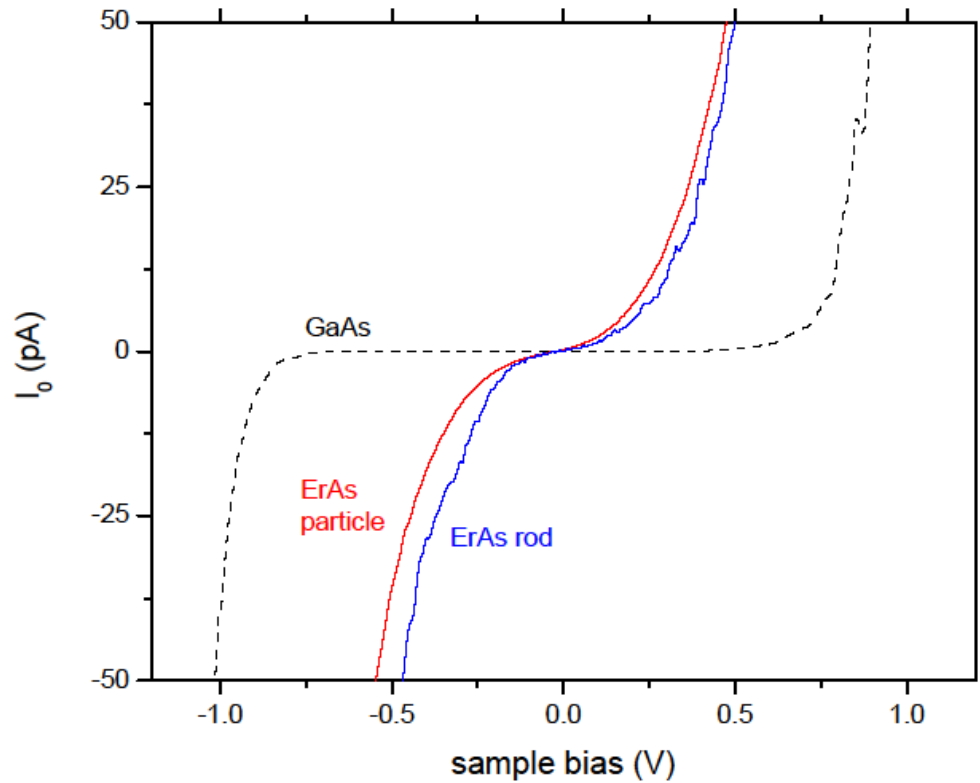

Figure 5. (Color online) Tunneling current spectroscopy for the GaAs matrix (dotted black), ErAs nanoparticles (red), and ErAs nanorods (blue). The nonzero slope at V=0 for both nanorods and nanoparticles suggests semimetallic behavior.